\documentclass[conference]{IEEEtran}
\ifCLASSINFOpdf
\else
\fi
\usepackage{booktabs}   
\usepackage{subcaption} 
\usepackage{xspace}
\usepackage{amsmath}
\usepackage{amsthm}
\usepackage{thmtools}
\usepackage{bussproofs}
\usepackage{syntax} 
\usepackage{enumitem}
\usepackage{color}
\usepackage{xcolor}
\usepackage{tikz}
\usepackage{pgfplots}
\usepackage{subcaption}
\usepackage{listings}

\usepackage{algorithm}
\usepackage[noend]{algpseudocode}
\usepackage{amssymb}
\usepackage{pifont}
\usepackage{multirow}
\usepackage{wasysym}
\newtheorem{definition}{Definition}
\newtheorem{example}{Example}

\newtheorem{vul}{Vulnerability}

\usepackage{url}
\usepackage{hyperref}

\begin{document}
%
\title{Precise Attack Synthesis for Smart Contracts}

\author{\IEEEauthorblockN{Yu Feng}
\IEEEauthorblockA{UC Santa Barbara\\
yufeng@cs.ucsb.edu}
\and
\IEEEauthorblockN{Emina Torlak}
\IEEEauthorblockA{University of Washington\\
emina@cs.washington.edu}
\and
\IEEEauthorblockN{Rastislav Bodik}
\IEEEauthorblockA{University of Washington\\
bodik@cs.washington.edu\\
}}


%


\maketitle



%
\IEEEpeerreviewmaketitle

\newcommand{\program}{P\xspace}
\newcommand{\abi}{I\xspace}
\newcommand{\query}{Q\xspace}
\newcommand{\lang}{\mathcal{LABIOMANCY}}
\newcommand{\toolname}{{\sc SmartScopy}\xspace}
\newcommand{\oyente}{{\sc Oyente}\xspace}
\newcommand{\rosette}{{\sc Rosette}\xspace}
\newcommand{\madmax}{{\sc Madmax}\xspace}
\newcommand{\contractfuzz}{{\sc ContractFuzzer}\xspace}
\newcommand{\pstate}{\Gamma}
\newcommand{\vulnerability}{\mathcal{V}}
\newcommand{\transition}{\mathcal{T}}
\newcommand{\contract}{\mathcal{C}}
\newcommand{\summary}{\mathcal{M}}
\newcommand{\pc}{\phi}
\newcommand{\store}{\sigma}
\newcommand{\ppath}{\pi}
\newcommand{\assert}{\alpha}
\newcommand{\todo}[1]{{\color{red}{#1}}}
\newcommand{\eval}[1]{{[\![#1]\!]}}
\newcommand{\sumi}[1]{{\overline{#1}}}
\newcommand{\geval}[1]{{[\![#1]\!]}}
\newcommand{\peval}[1]{{[\![#1]\!]_\pstate}}
\newcommand{\reach}{\rightsquigarrow}

\newcommand{\hypo}{\mathcal{G}}
\newcommand{\sketch}{\mathcal{S}}
\newcommand{\victim}{V}
\newcommand{\sample}{\Delta}
\newcommand{\comps}{\Upsilon}
\newcommand{\ex}{\mathcal{E}}
\newcommand{\worklist}{\mathcal{W}}
\newcommand{\prog}{\mathcal{P}}
\newcommand{\comp}{\mathcal{C}}
\newcommand{\tab}{\mathcal{T}}
\newcommand{\cmark}{\ding{51}}%
\newcommand{\xmark}{\ding{55}}%

\lstset{basicstyle=\footnotesize\ttfamily,breaklines=true,numbers=left,stepnumber=1}
\lstset{frame=bottomline}

\begin{abstract}
Smart contracts are programs running on top of blockchain platforms. They
interact with each other through well-defined interfaces to perform financial
transactions in a distributed system with no trusted third parties. But these
interfaces also provide a favorable setting for attackers, who can exploit
security vulnerabilities in smart contracts to achieve financial gain. 

This paper presents \toolname, a system for automatic synthesis of adversarial
contracts that identify and exploit vulnerabilities in a victim smart contract.
Our tool explores the space of \emph{attack programs} based on the Application
Binary Interface (ABI) specification of a victim smart contract in the Ethereum
ecosystem. To make the synthesis tractable, we introduce \emph{summary-based
symbolic evaluation}, which significantly reduces the number of instructions
that our synthesizer needs to evaluate symbolically, without compromising the
precision of the vulnerability query. Building on the summary-based symbolic
evaluation, \toolname further introduces a novel approach for partitioning the
synthesis search space for parallel exploration, as well as a lightweight deduction
technique that can prune infeasible candidates earlier. We encoded common
vulnerabilities of smart contracts in our query language, and evaluated
\toolname on the entire data set from etherscan with $>$25K smart contracts. Our
experiments demonstrate the benefits of summary-based symbolic evaluation and
show that \toolname outperforms two state-of-the-art smart contracts analyzers,
\oyente and \contractfuzz, in terms of running time, precision, and soundness.
Furthermore, running on recent popular smart contracts, \toolname uncovers
20 vulnerable smart contracts that contain the recent BatchOverflow vulnerability and cannot be precisely detected by existing tools. 
\end{abstract}
\section{Introduction}\label{sec:intro}
Smart contracts are programs running on top of blockchain platforms such as
Bitcoin~\cite{bitcoin} and Ethereum~\cite{ethereum}.  They have been receiving
much attention due to the capability to perform effective financial transactions
in a distributed system without the intervention from trusted third parties
(e.g., banks). A smart contract is written in a high-level programming language
(e.g., Solidity~\cite{solidity}), and it is typically comprised of a unique
address, persistent storage holding a certain amount of cryptocurrency (i.e.,
Ether in Ethereum), and a set of functions that manipulate the persistent
storage to fulfill credible transactions without trusted parties. For
contract-to-contract interaction, some functions are public and callable by
other contracts. Thanks to the expressiveness afforded by the high-level
programming languages and the security guarantees from the underlying consensus
protocol, smart contracts have shown many attractive use cases, and their number
has skyrocketed, with over 45 million~\cite{etherscan} instances covering
financial products, online gaming, real estate~\cite{case1}, shipping, and
logistics~\cite{case2}.

Because all smart contracts deployed on a blockchain are freely accessible
through their public methods, any functional bugs or vulnerabilities inside the
contracts can lead to disastrous losses, as demonstrated by recent
attacks~\cite{attack1,attack2,attack3,attack4}. For instance, the code in
Figure~\ref{fig:intro-example} illustrates a reentrancy
vulnerability exploited in the notorious DAO attack~\cite{attack1}. When the
victim program issues a money transaction (line 9 in Figure~\ref{fig:dao-vic})
to the attacker, it implicitly triggers the attacker's callback method (line 3
in Figure~\ref{fig:dao-attack}), which invokes the victim's method again to make
another transaction without updating the victim's balance. The DAO attack led to
a financial loss of \$150M in 2016. To make things worse, smart contracts are
immutable---once they are deployed, fixing their bugs is extremely difficult due
to the design of the consensus protocol.\looseness=-1

Improving robustness of smart contracts is thus a pressing practical problem.
It is also an active area of research, with several contract analysis 
tools~\cite{oyente,securify,contractfuzzer,madmax,zeus,teether} developed in the
past few years. However, these tools either soundly overapproximate the
execution of smart contracts and report warnings~\cite{securify,madmax} that
cannot be exploited in reality, or they precisely
enumerate~\cite{teether,contractfuzzer,oyente} \emph{concrete traces} of smart
contracts, so cannot scale to analyze large programs.


\begin{figure}
    \begin{subfigure}[b]{0.4\textwidth}
    \begin{lstlisting}
contract Victim {
  private userBalances;

  function withdraw() public {
    uint amount = balances[msg.sender];
    //call withdrawBalance again
    msg.sender.call.value(amount)(); 
    balances[msg.sender] = 0;
  }
}
    \end{lstlisting}
    \caption{The Vulnerable Program}
      \label{fig:dao-vic}
    \end{subfigure} \\
    \begin{subfigure}[b]{0.4\textwidth}
    \begin{lstlisting}      
  contract Attacker {
    ...
    function () payable {
      Victim v;
      v.withdraw();
    }
  }
\end{lstlisting}
    \caption{The Attack Program}
      \label{fig:dao-attack}
    \end{subfigure}
\caption{Reentrancy Attack \label{fig:intro-example}}
\end{figure}

\begin{figure*}
  \centering
  \includegraphics[width=\textwidth]{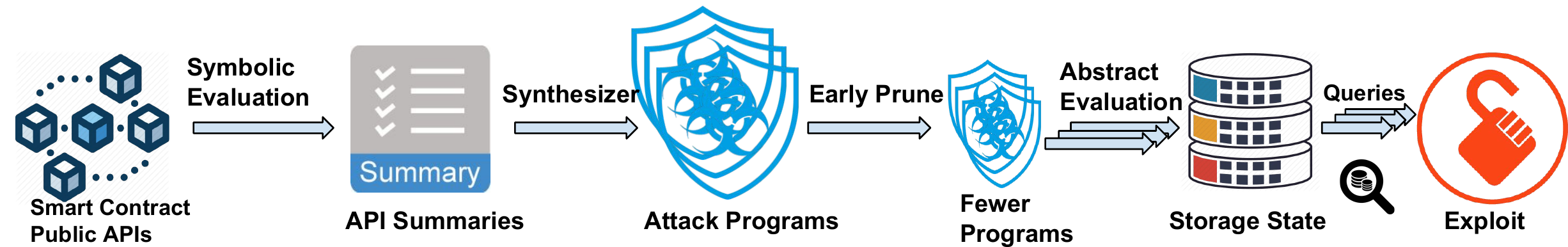}
\caption{Overview of \toolname}
\label{fig:overview}
\end{figure*}

This paper presents \toolname, a tool that uses \emph{program synthesis} to
automatically generate adversarial smart contracts (i.e., attack programs),
which exploit common vulnerabilities in victim contracts. To use our tool, a
security analyst expresses a target vulnerability query (e.g., the reentrancy
vulnerability from the DAO attack) as a declarative specification. Then,
\toolname \emph{synthesizes} an attack program that exploits the victim's public
interface to satisfy the vulnerability query. Given this problem, a naive
approach is to enumerate all possible candidate programs and then symbolically
evaluate each of them to check if it satisfies the query. While precise, the
naive approach fails to scale to realistic contracts. To tackle this challenge,
we employ a novel \emph{summary-based symbolic evaluation}, which enables
\toolname to both find real attacks and scale to large programs. 

Fig~\ref{fig:overview} shows an overview of our approach. Given the public
methods provided by the Application Binary Interface (ABI) of a smart contract,
our system first \emph{symbolically} evaluates each method and generates a
summary that soundly records the method's side-effects on the storage as well as
other global state of the Blockchain. Even with the summaries, the search space
is still too large for brute-force enumeration. To address this issue, we
partition the search space by case splitting on the range of symbolic variables,
which allows us to simultaneously explore multiple attack programs using an
SMT-based symbolic evaluation engine~\cite{rosette}. \toolname further reduces
the search space by pruning infeasible candidates early, using their
\emph{abstract semantics}. After that, our tool symbolically evaluates each
remaining candidate to check if any of them satisfies the vulnerability query.
If so, the candidate is returned as a potential exploit.

We have evaluated \toolname on the entire data set ($>$25K) from
etherscan~\cite{etherscan} and shown that our tool is expressive, efficient, and
effective. \toolname's query specification language is expressive in that it is
rich enough encode common vulnerabilities found in the literature (such as the
Reentrancy attack~\cite{attack1}, Time manipulation~\cite{attack-time}, and
malicious access control~\cite{teether}), Security Best
Practices~\cite{best-practice}, as well as the recent batchOverflow
Bug~\cite{attack-int} (CVE-2018–10299), which allows the attacker to create an
arbitrary amount of cryptocurrency. \toolname is efficient: on average it takes
only 8 seconds to analyze a smart contract from etherscan, which is an
order of magnitude faster than \oyente~\cite{oyente} and two orders of magnitude
faster than \contractfuzz~\cite{contractfuzzer}. \toolname is also effective in
that it significantly outperforms two state-of-the-art smart contracts
analyzers, namely, \oyente and \contractfuzz, in terms of false positive and
false negative rates. Furthermore, running on recent popular smart contracts,
\toolname uncovers 20 vulnerable contracts that contain the BatchOverflow vulnerability and cannot be precisely detected by
existing tools. 

In summary, this paper makes the following key contributions:\looseness=-1
\begin{itemize}
\item We formalize the problem of exploit generation as a program synthesis problem 
and provide a way of expressing common vulnerabilities in smart contracts as declarative
specifications (Section~\ref{sec:vul}).
\item We propose a summary-based symbolic evaluation technique that significantly 
reduces the number of instructions that \toolname has to execute (Section~\ref{sec:sum}).
\item We develop an efficient attack synthesizer based on the summary-based symbolic 
evaluation, which incorporates a novel combination of search space partitioning, parallel
symbolic execution, and early pruning based on the abstract semantics of
candidate programs (Section~\ref{sec:parallel}).
\item We perform a systematic evaluation of \toolname on the entire data set
from etherscan. Our experiments demonstrate the substantial benefits of our
technique and show that \toolname outperforms two state-of-the-art smart
contracts analyzers in terms of running time, precision, and soundness
(Section~\ref{sec:eval}).
\end{itemize}

\section{Background}\label{sec:background}
This section briefly reviews the background on blockchains and smart contracts.

\subsection{Blockchain and Ethereum}
Blockchain, invented by Satoshi Nakamoto in 2008, is a distributed public ledger that stores transactions between different parties. A blockchain is comprised of a growing list of blocks, each of which contains the hash of the previous block, a timestamp when the block is appended, and transaction value. Due to the decentralized consensus protocol, each block is inherently resistant to modification once it is created.\looseness=-1 

While Satoshi's original blockchain proposes a peer to peer e-cash system that offers secure transactions, the Ethereum~\cite{yellowpaper} blockchain provides a more powerful distributed computing platform that can execute custom code in the form of \emph{smart contracts}. In addition to the crypto tokens (i.e., Ether) that are transferred among parties during a transaction, Ethereum also implements a \emph{gas} scheme (explained in Section~\ref{sec:abi}) to incentivize \emph{miners} who perform the computationally expensive creation of new blocks. 

\subsection{Smart Contract}
Smart contracts are programs that are stored and executed on the blockchain. They are created through the transaction system on the blockchain and are immutable once deployed. Each smart contract is associated with a unique 256-bit address; a private persistent storage; a certain amount of cryptocurrency, denoted by balance (i.e., Ether in Ethereum) held by the contract; and a piece of executable code that fulfills complex computations to manipulate the storage and balance. The code is typically written in a high-level Turing-complete programming language such as Serpent~\cite{serpent}, Vyper~\cite{vyper}, and Solidity~\cite{solidity}, and then compiled to the Ethereum Virtual Machine (EVM) bytecode~\cite{yellowpaper}, a low-level stack-based language. 
For instance, Figure~\ref{fig:attack-vic} shows a smart contract written in the Solidity programming 
language~\cite{solidity}.

\subsection{ABI and Transactions}\label{sec:abi}
In the Ethereum ecosystem, Smart Contracts communicate with each other using the Contract Application Binary Interface (ABI), which defines the signatures of public functions provided by the hosted contract. While ABI offers a flexible mechanism for communication, it also creates an attack surface for exploits that use  the ABI of a given smart contract. We will elaborate on this in the following section.
For instance, Figure~\ref{fig:attack-abi} shows the ABI for the smart contract in Figure~\ref{fig:attack-vic}.
\begin{table}[]
\centering
\begin{tabular}{|l|l|}
\hline
...         & ...                                                                \\ \hline
From:       & 0x7d5c8c59837357e541bc7d87dee53fcbba55ba65                         \\ \hline
To:         & 0x8811fffcfc266844e8c36418389f7cda76c77ab7 \\ \hline
Value:      & 0.05 Ether                                                         \\ \hline
Gas Limit:  & 31602                                                              \\ \hline
Input Data: & 0x687474703a2f2f6c6f63616c686f73743a38353435                       \\ \hline
\end{tabular}
\caption{A sample transaction~\cite{sample-trans} obtained from Etherscan}
\label{fig:trans-sample}
\end{table}

All interactions between smart contracts are fulfilled 
by transactions. Table~\ref{fig:trans-sample} shows a sample transaction obtained 
from Etherscan. Here, the important fields are \texttt{From},
\texttt{To}, \texttt{Value}, \texttt{Gas Limit}, and \texttt{Input Data}. In 
particular, \texttt{From} and \texttt{To} represent the sender and recipient, respectively.
\texttt{Value} denotes the amount transferred from one smart contract to another. 
\texttt{Input Data} contains the function's signature (obtained from the ABI) and its arguments. 
Finally, the \texttt{Gas Limit} field specifies the amount of cryptocurrency which a miner 
gets for conveying the transaction. The Ethereum protocol~\cite{yellowpaper} defines
the gas cost for each bytecode instruction. For instance, an integer division operation 
costs 5 units of gas while a store operation on the storage can cost up to 20000.
As we will see in Section~\ref{sec:eval}, 
the gas mechanism plays a key role in several different types of vulnerabilities.

\subsection{Threat Model}

To synthesize an adversarial contract,  we assume that we can obtain the victim 
contract's bytecode and the ABI specifying its public methods. To confirm 
an adversarial contract is indeed an exploit, we must also be able to invoke public methods 
by submitting transactions over the Ethereum Blockchain. 
These requirements are easy to satisfy in practice.

\section{Overview}\label{sec:overview}
In this section, we give an overview of our approach with the aid of a motivating example.

\subsection{Smart Contract Vulnerabilities}\label{sec:overview-vul}

A security analyst can specify various types of vulnerabilities that may appear
in a smart contract. 
For instance, a \emph{Reentrancy vulnerability}~\cite{attack1} 
occurs when an attacker's previous invocation is allowed to make new calls to the 
victim contract before the previous execution is complete. This means that if the 
call involves money transactions, the attacker can repeatedly 
trigger many transactions until the current procedure runs out of gas. A
\emph{Timestamp dependence vulnerability}~\cite{attack-time}, on the other hand, 
happens when a transaction relies on a certain timestamp, which allows malicious miners to gain advantage by choosing a suitable timestamp. 

This section uses the most recent \emph{BatchOverflow 
Vulnerability (CVE-2018–10299)}~\cite{attack-int} as a motivating example.
Exploits due to this vulnerability have resulted in the creation of 
trillions of \emph{invalid} Ethereum Tokens in 2018~\cite{batch-news}, causing major exchanges to temporary halt until all tokens could be reassessed.
As shown in Fig~\ref{fig:attack-vic}, the \texttt{batchTransfer} function 
performs a multiplication that can overflow 256 bits, 
which results in a small value that passes the check at line 12 and 
further transfers a lage amount of tokens to the attacker (line 18).\looseness=-1

\toolname's specifications are assertions expressed in the Racket 
language~\cite{racket}. In particular, the \emph{BatchOverflow Vulnerability}
can be expressed as follows:
\begin{lstlisting}      
$\exists\ arg_0, arg_1, r_1, r_2, r_3, call$ 
(&& (= $r_3$ ($\otimes$ $r_1$ $r_2$)) 
    (> $\geval{r_2}$ $\geval{r_3}$)
    (interfere? $r_2$ call.value) 
    (interfere? $arg_0$ call.addr) 
    (interfere? $arg_1$ call.value)) 
where $\otimes \in${+,$\times$}
\end{lstlisting}
\begin{figure}
  \centering
  \includegraphics[scale=0.4]{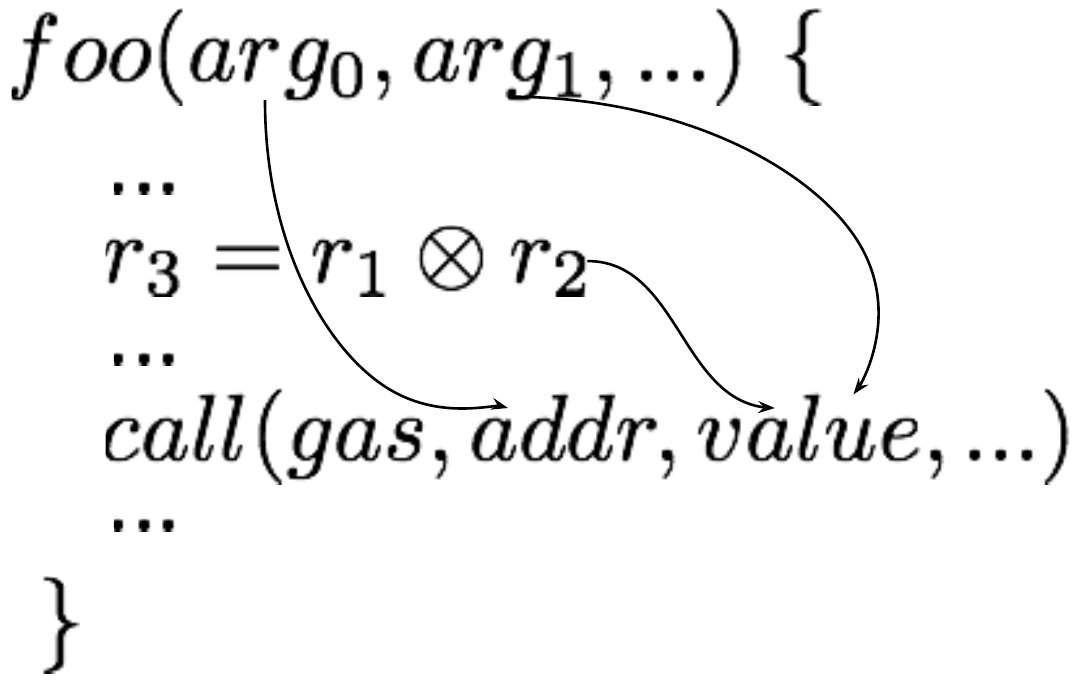}
\caption{The key pattern of the BatchOverflow Vulnerability}
\label{fig:batchcode}
\end{figure}
We also visualize this vulnerability pattern in Fig~\ref{fig:batchcode}. 
Here, $arg_i$, $r_j$, and \texttt{call} represent function arguments, 
registers, and the \texttt{CALL} instruction (to perform a transaction in Solidity), respectively. 
We use $\geval{r_i}$ to denote the value 
(either concrete or symbolic) in the register $r_i$.
The \texttt{interfere?} function, which is defined
in section~\ref{sec:problem}, checks the interference between two 
expressions. The interference~\cite{non-interfere} 
(denoted by an arrow in Fig~\ref{fig:batchcode}) in our system precisely
captures the data- and control-dependency. For instance,   
the vulnerability states that, there exists a \texttt{CALL} instruction
for which the beneficiary (i.e., recipient's address) and value are controlled by the attacker (line 5, 6).
Furthermore, the transaction's value is influenced by a register (line 4) used in an arithmetic operation that overflows (line 2,3).

Once a security analyst expresses the Batchoverflow vulnerability, the next step  
is to construct an attack to confirm that the vulnerability indeed exists in the victim contract. 
Doing so manually is challenging, however, because the analyst has to understand the 
semantics of the smart contract and simulate all possible interactions that 
an attacker may perform.  
As a result, the analysis process is both tedious and error-pone.
\begin{figure}
    \begin{subfigure}[b]{0.4\textwidth}
    \begin{lstlisting}[escapechar=@]  
contract PausableToken {
bool flag = false; 

function makeFlag(bool fg) { 
 flag = fg; 
} 

function batchTransfer(address[] _receivers, uint256 _value) {
    uint cnt = _receivers.length;
    uint256 amount = @\textbf{uint256(cnt) * _value}@;
    require(@\textbf{flag}@);
    require(balances[msg.sender] @\textbf{>= amount)}@;

    balances[msg.sender] = 
      balances[msg.sender].sub(amount);
    for (uint i = 0; i < cnt; i++) {
      address recv = _receivers[i];
      balances[recv] = 
        balances[recv].add(_value);
      @\textbf{Transfer(msg.sender, recv, _value)}@;
    }
    return true;
  }
}
\end{lstlisting}
    \caption{The Vulnerable Program}
      \label{fig:attack-vic}
    \end{subfigure} \\
    \begin{subfigure}[b]{0.4\textwidth}
    \begin{lstlisting}      
  contract Attacker {
    ...
    function exploit() {
      VulContract v;
      v.makeFlag(true);
      v.batchTransfer([0x123, 0x456], $2^{256} - 1$);
    }
  }
\end{lstlisting}
    \caption{An Attack Program}
      \label{fig:attack-code}
    \end{subfigure} \\
    \begin{subfigure}[b]{0.4\textwidth}
    \begin{lstlisting} 
 {
  "inputs": [
  {"name": "_receivers", "type": "address[]"},
  {"name": "_value", "type": "uint256"}
  ],
  "name": "batchTransfer", "type": "function"
  ...
},{ 
 "inputs": [{"name": "fg", "type": "bool"}],
 "name": "makeFlag", "type": "function"}
  \end{lstlisting}
      \caption{Contract Application Binary Interface (ABI) for the vulnerable contract in Fig~\ref{fig:attack-vic}}
      \label{fig:attack-abi}
    \end{subfigure}
\caption{A running example to show the BatchOverFlow Vulnerability\label{fig:overview-example}}
\end{figure}

\subsection{\toolname}

\toolname helps automate this process by searching for attacks that 
exploit a given vulnerability in a victim contract. As shown in Fig~\ref{fig:overview}, the tool takes as input a 
potential vulnerability $\vulnerability$ expressed as declarative specifications. If  
 $\vulnerability$ exists in the victim contract, \toolname {automatically synthesize} an 
\emph{attack program} that exploits $\vulnerability$. In practice, an attacker  
typically interacts with a vulnerable contract through its public methods defined 
in the ABI. Therefore, our goal is to construct an attack program that exploits the victim's ABI and that contains at least one concrete
trace where $\vulnerability$ holds. 

To achieve this goal, \toolname  models 
the executions of a smart contract as \emph{state transitions} over registers, memory, and storage. 
The vulnerability $\vulnerability$ 
is expressed in Racket as a boolean predicate over these state transitions. 
The technical challenge addressed by \toolname is to efficiently search for an attack program where $\vulnerability$ 
holds. 

To illustrate the difficulty of this task, consider the problem of synthesizing an attack 
program that exploits 
the BatchOverFlow vulnerability in Fig~\ref{fig:overview-example}. 
The attack program performs a complex three-step 
interaction with the victim contract.  First, the attacker must set the storage variable 
\texttt{flag} to \texttt{true} to pass the check at line 11. Next, it needs to 
assign a large number to \texttt{_value} that leads to an overflow at line 10. 
Finally, it specifies the attacker's address as the beneficiary of the transaction (line 18). 
Synthesizing this attack program involves discovering which methods to call, in what order, and with what arguments. 


To find the desired attack program, it is not feasible to brute-force generate 
all possible \emph{concrete programs} and explore the space of 
their \emph{concrete traces}. As we elaborate in Section~\ref{sec:overview-vul},
the search space is exponential to the size of the attack program as well as 
its the number of branches.

To address this challenge, Section~\ref{sec:sum} proposes a novel summary-based 
symbolic evaluation technique that significantly reduces the number of 
instructions in the victim contract that \toolname has to execute symbolically (Section~\ref{sec:sum}). Intuitively,
our summary-based symbolic evaluation enables \toolname to only \emph{preserve state transitions} that 
are persistent across different transactions and are \emph{sufficient} to answer
the vulnerability query.\looseness=-1

Even with our summary-based symbolic evaluation, the search space of attack candidates 
is still too large for brute-force search.
To further improve the performance, Section~\ref{sec:impl} 
introduces three optimizations. First, instead of exploring the space of concrete
programs, we leverage \rosette~\cite{rosette} to explore the \emph{symbolic programs} (Section~\ref{sec:rosette}). Second, instead of eagerly explore the space
of symbolic programs,  we design a simple but effective \emph{early pruning}
strategy that allows \toolname to prune \emph{infeasible} symbolic candidates 
before executing them (Section~\ref{sec:prune}). Finally, instead of executing 
each symbolic program \emph{sequentially}, we partition the search space by case splitting 
on the range of symbolic variables, which enables \toolname to simultaneously 
explore multiple symbolic candidates (Section~\ref{sec:parallel}).

\section{Problem Formulation}\label{sec:problem}

This section formalizes the semantics of smart contracts, shows how to express
Smart Contract Vulnerabilities, and defines what it means for a vulnerability to
appear in a smart contract.\looseness=-1

\subsection{Smart Contract Language}\label{sec:lang}

Figure~\ref{fig:grammar} shows the core features of our intermediate language
for smart contracts. This language is a superset of the EVM language. It includes
standard EVM bytecode instructions such as assignment (\texttt{x := e}), memory operations
(\texttt{mstore,mload}), storage operations (\texttt{sstore,sload}), hash
operation (\texttt{sha3}), sequential composition (\texttt{$s_1;s_2$}),
conditional (\texttt{jumpi}) and unconditional jump (\texttt{jump}). It also
includes the EVM instructions specific to smart contracts: \texttt{call}
transfers the balance from the current contract to a recipient whose address is
specified as the argument, \texttt{balance} accesses the current account
balance, and \texttt{selfdestruct} terminates a contract and transfers its
balance to a given address. Finally, our language extends EVM with features that
facilitate symbolic evaluation, including \emph{symbolic variables} (introduced
by \texttt{def-sym}) and \emph{symbolic expressions} (obtained by operating on
symbolic variables) whose concrete values will be determined by an off-the-shelf
SMT solver~\cite{NiemetzPreinerBiere-JSAT15}.

\begin{figure}[!t]
\begin{grammar}
  <var>  ::= \texttt{def-sym} (id $\tau$) \\
  ($\tau\in\{\textbf{boolean}, \textbf{number}\}$)

  <pc>   ::= <const> | <var>

  <expr> ::= <const> | <var> | <expr> $\oplus$ <expr> \\
  ($\oplus\in\{+, -, \times, /, \vee, \wedge, ...\}$)

  <stmt> ::=  <var> := <expr> 
   \alt <var> := \textbf{mload} <var> | \textbf{mstore} <var> <var>
   \alt <var> := \textbf{sload} <var> | \textbf{sstore} <var> <var>
   \alt <var> := \{\textbf{balance}, \textbf{gas}, \textbf{address} \}

  <stmts> ::= <stmt> | <stmt>; <stmts> | \textbf{sha3} <var> <var>
  \alt \textbf{jumpI} <pc> <expr> | \textbf{jump} <pc> | \textbf{no-op}
  \alt \textbf{call} <var> <var> <var> | \textbf{selfdestruct} <var>

  <param> ::= <var> 

  <params> ::= <param> | <param>, <params>

  <prog> ::= $\lambda$<params>. <stmts> 
\end{grammar}
\caption{Intermediate Language for Smart Contract}
\label{fig:grammar}
\end{figure}

\begin{figure}[!t]

\begin{prooftree}
\AxiomC{$s = \textbf{no-op}$}
\LeftLabel{\texttt{no-op}}
\UnaryInfC{$\pstate \vdash s: \pstate'[\text{pc++}]$}
\end{prooftree}
\vspace{0.05in}

\begin{prooftree}
\AxiomC{$s=(\textbf{jumpi } d \ e)$}
\noLine
\UnaryInfC{$\pstate \vdash e: v$}
\AxiomC{$i=(v=0)?(pc+1):d$}
\noLine
\UnaryInfC{$\pstate' = \pstate[x\gets v,pc\gets i]$}
\LeftLabel{\texttt{jmp}}
\BinaryInfC{$\pstate \vdash s: \pstate'$}
\end{prooftree}
\vspace{0.05in}

\begin{prooftree}
\AxiomC{$s = (param := \textbf{def-sym}(e,\tau))$}
\AxiomC{$v = |(e,\tau)|$}
\noLine
\UnaryInfC{$\pstate' = \pstate[param\gets v]$}
\LeftLabel{\texttt{sym}}
\BinaryInfC{$\pstate \vdash s: \pstate', v$}
\end{prooftree}
\vspace{0.05in}

\begin{prooftree}
\AxiomC{$s = (x := e)$}
\AxiomC{$\pstate \vdash e: v$}
\AxiomC{$\pstate' = \pstate[x\gets v]$}
\LeftLabel{\texttt{assign}}
\TrinaryInfC{$\pstate \vdash s: \pstate'[\text{pc++}], v$}
\end{prooftree}
\vspace{0.05in}

\begin{prooftree}
\AxiomC{$s = (x:=e_1 \oplus e_2)(\oplus \in \{+,-,/,\times\})$ }
\noLine
\UnaryInfC{$\pstate \vdash e_1: v_1$}
\noLine
\UnaryInfC{$\pstate \vdash e_2: v_2$}
\AxiomC{$v=v_1 \oplus v_2$}
\noLine
\UnaryInfC{$\pstate'[x\gets v]$}
\LeftLabel{\texttt{biop}}
\BinaryInfC{$\pstate \vdash s: \pstate'[\text{pc++}]$}
\end{prooftree}
\vspace{0.05in}

\begin{prooftree}
\AxiomC{$s = s_1;s_2$ }
\AxiomC{$\pstate \vdash s_1: \pstate_1, v_1$}
\noLine
\UnaryInfC{$\pstate_1\vdash s_2: \pstate_2, v_2$}
\LeftLabel{\texttt{seq}}
\BinaryInfC{$\pstate \vdash s: \pstate_2, v_2$}
\end{prooftree}
\vspace{0.05in}

\begin{prooftree}
\AxiomC{$s = (x:=\textbf{sload } e)$ }
\noLine
\UnaryInfC{$\pstate \vdash e: \mu$}
\AxiomC{$\pstate \vdash \mu: v$}
\noLine
\UnaryInfC{$\pstate'[x\gets v]$}
\LeftLabel{\texttt{sload}}
\BinaryInfC{$\pstate \vdash s: \pstate'[\text{pc++}],v$}
\end{prooftree}
\vspace{0.05in}

\begin{prooftree}
\AxiomC{$s = \textbf{sstore } \mu \ e$ }
\noLine
\UnaryInfC{$\pstate \vdash e: v$}
\AxiomC{$\pstate'[\mu\gets v]$}
\LeftLabel{\texttt{sstore}}
\BinaryInfC{$\pstate \vdash s: \pstate'[\text{pc++}]$}
\end{prooftree}
\vspace{0.05in}

\begin{prooftree}
\AxiomC{$s = (r_l:=\textbf{call }(e_1 \ e_2 \ e_3))$ }
\noLine
\UnaryInfC{$\pstate \vdash e_1: v_1$}
\noLine
\UnaryInfC{$\pstate \vdash e_2: v_2$}
\AxiomC{$\pstate \vdash e_3: v_3$}
\noLine
\UnaryInfC{$v=\textbf{call}_i(v_1,v_2,v_3)$}
\noLine
\UnaryInfC{$\pstate'[r_l\gets v]$}
\LeftLabel{\texttt{call}}
\BinaryInfC{$\pstate \vdash s: \pstate'[\text{pc++}], r_l$}
\end{prooftree}
\vspace{0.05in}

\begin{prooftree}
\AxiomC{$s = (x:= \textbf{sha3 }m \ e)$ }
\noLine
\UnaryInfC{$\pstate \vdash m: v_1$}
\noLine
\AxiomC{$\pstate \vdash e: v_2$}	
\noLine
\UnaryInfC{$v=\textbf{sha3}_i(v_1,v_2)$}
\noLine
\UnaryInfC{$\pstate'[x\gets v]$}
\LeftLabel{\texttt{sha}}
\BinaryInfC{$\pstate \vdash s: \pstate'[\text{pc++}],v$}
\end{prooftree}
\caption{Operational semantics for our language 
in Figure~\ref{fig:grammar}}
\label{fig:sem}
\end{figure}

We define the semantics of the language operationally, as 
shown in Figure~\ref{fig:sem}. The meaning of each statement 
is given by a \emph{state transition} rule that specifies the statement's 
effect on the \emph{program state}. We define states and transitions as follows. 
\begin{definition}{{(\bf Program State)}}
The \emph{Program State} $\pstate$ consists of a stack $E$, memory $M$,
 persistent storage $S$, global properties (e.g., balance, address, timestamp)
 of a smart contract, and the program counter \texttt{pc}. We use $e_i$, $m_i$,
 and $\mu_i$ to denote variables from the stack, memory, and storage,
 respectively. 
\end{definition}
\noindent A program state also includes a model of the gas system in EVM, but we
omit this part of the semantics to simplify the presentation. If a state maps a
variable to a symbolic expression, we call it an \emph{abstract state}.  
\begin{definition}{{(\bf State Transition over statement $s$)}}
  A \emph{State Transition} $\transition$ over a statement $s$ is
  denoted by a judgment of the form $\pstate \vdash s: \pstate', v$. 
  The meaning of this judgment is the following: assuming we successfully execute $s$ under program 
  state $\pstate$, it will result in value $v$ and the new state is $\pstate'$. 
  We use $\pstate \vdash s: \bot$ to indicate failure.
\end{definition}

Most of the rules in Figure~\ref{fig:sem} specify the standard semantics of EVM 
instructions. For example, the \texttt{biop} rule describes the meaning of binary operations: 
it first looks up the values (concrete or symbolic) of the 
operands in the current program state $\pstate$, applies the binary operator to those
values (i.e., $v_1, v_2$), and then binds the result to the target variable, 
increases the program counter, and produces a new program state $\pstate'$. 
The  \texttt{sstore}, \texttt{sload}, \texttt{jmp}, and \texttt{seq} rules are also standard.

The  \texttt{sym},  \texttt{sha3}, and  \texttt{call} rules, on the other hand, are 
tailored for (efficient) symbolic evaluation. 
The \texttt{sym} rule introduces symbolic values into the program state. 
The construct $|(e,\tau)|$ denotes a fresh symbolic variable $e$ of type $\tau$,
which is bound to the \texttt{def-sym} parameter in the new program state
$\pstate'$. Here, we do not increase the program counter as the symbolic binding
is not an EVM instruction.
The \texttt{sha3} and \texttt{call} instructions are part of EVM, but we overapproximate their 
semantics with  \emph{uninterpreted functions} to produce more tractable vulnerability queries. 

The standard semantics of the \texttt{sha3} instruction is to obtain a memory 
location by hashing a memory address and offset. 
However, applying hashing functions to symbolic arguments results in hard-to-solve queries. 
The \texttt{sha3} rule therefore uses an uninterpreted function, denoted by $\texttt{sha3}_i$,  
to model the original hash function. 

As mentioned earlier, the \texttt{call} instruction is used to initiate 
a transaction with another contract, whose address is specified as an argument.
The \texttt{call} rule uses an 
uninterpreted function, denoted by $\texttt{call}_i$, to model the effect of the \texttt{call} instruction. 
Note that the rule also records the return value of each \texttt{call} using a special variable 
$r_l$ in $\pstate$, where $l$ is the location of the \texttt{call} command. 
This handling of \texttt{call} instructions is key to our summary-based symbolic evaluation, as explained 
in Section~\ref{sec:sum}.\looseness=-1


\begin{figure*}
    \centering
    \begin{subfigure}[b]{0.4\textwidth}
\begin{lstlisting}[language=Java]
pragma solidity ^0.4.23;
contract EubChainIco is PausableToken {
  ...
  function vestedTransfer(address _to, uint256 _amount) {
      ...
      require(_amount > 0); 
      vesting.amount = _amount.sub(1);
      transfer(msg.sender, _to, vesting.amount);

      uint256 v1 = _amount - 15;
      uint256 wei = v1;
      uint t1 = vesting.startTime;

      emit VestTransfer(msg.sender, _to, wei, t1, _);
      ...
  }
}
\end{lstlisting}
      \caption{A Smart Contract written in Solidity.}
      \label{fig:code-sum}
    \end{subfigure}
    ~ 
    \begin{subfigure}[b]{0.4\textwidth}
      \begin{lstlisting}[title={(b) Symbolic Interpretation},basewidth=0.5em]
        assert(_amount > 0);
        r1 (*\textbf{:=}*) _amount - 1;
        (*\textbf{sstore}*)(vesting.amount, _amount - 1);
        (*\textbf{call}*)(msg.sender, _to, _amount - 1);

        r2 (*\textbf{:=}*) amount - 15;
        r3 (*\textbf{:=}*) amount - 15;
        r4 (*\textbf{:=}*) (*\textbf{sload}*)(vesting.startTime);
        (*\textbf{no-op}*);
      \end{lstlisting}
      \begin{lstlisting}[title={(c) Summary Extraction},basewidth=0.5em]
        $\sumi{sstore}$(vesting.amount, _amount - 1) (*\textbf{[_amount>0]}*);
        $\sumi{call}$(msg.sender, _to, _amount - 1) (*\textbf{[_amount>0]}*);
      \end{lstlisting}

      \begin{lstlisting}[title={(d) Abstract Interpretation},basewidth=0.5em]
        if (_amount > 0)
          (*\textbf{sstore}*)(vesting.amount, _amount - 1);
        if (_amount > 0)
          (*\textbf{call}*)(msg.sender, _to, _amount - 1);
      \end{lstlisting}
    \end{subfigure}
    \caption{From Symbolic Interpretation to Summary-based Abstract Interpretation}
    \label{fig:sum-interp}
\end{figure*}

\begin{example}
Figure~\ref{fig:code-sum} shows a smart contract written in Solidity. To analyze this 
contract, our system first translates it to the program in Figure~\ref{fig:sum-interp}b,
using the intermediate language in Figure~\ref{fig:grammar}. The resulting program is then 
evaluated symbolically using the operational semantics in 
Figure~\ref{fig:sem}. For instance, after executing the statement at line 2 in 
Figure~\ref{fig:sum-interp}b, register \texttt{r1} holds a symbolic value represented 
by \texttt{_amount - 1}. On the other hand, since \toolname does not model the event 
system in Solidity, we simply turn all its corresponding instructions (e.g., line 14 in Figure~\ref{fig:code-sum}) into \texttt{no-op}.
\end{example}


\subsection{Smart Contract Vulnerabilities}\label{sec:vul}

Having defined the meaning of smart contracts, we now 
describe how to formally express smart contract vulnerabilities and
what it means for a vulnerability to appear in a program.

\begin{definition}{{(\bf Vulnerability)}}
A \emph{Vulnerability} $\vulnerability$ is a predicate over a set of 
variables $V$ in the program state. 
A vulnerability $\vulnerability$ appears in the program $P$ 
if the execution of $P$ can reach a program state $\pstate'$ that satisfies $\vulnerability$:
\[
    \pstate' \models \vulnerability
\]
\end{definition}

The rest of this section introduces a few representative vulnerabilities,
and shows how they are encoded as formulas in \toolname.
But first, we introduce an auxiliary function \texttt{interfere?} 
which will be used by several vulnerabilities.

\begin{definition}{{(\bf Interference)}}
A symbolic variable $v$ interferes with a symbolic expression $e$ if they satisfy
the following constraint: 
\[
    \exists v_0,v_1. \ e[v_0/v] \neq e[v_1/v] \land (v_0 \neq v_1)
\]
\end{definition}
\noindent Intuitively, changing $v$'s value will also affect $e$'s output,
which is denoted as ``(interfere? $v$ $e$)". Interference precisely captures 
the data- and control-dependencies between two expressions and turns out to be 
the \emph{necessary condition} of many exploits.

\medskip
Section~\ref{sec:overview} describes the BatchOverflow vulnerability, 
which enables an attacker to perform a multiplication that 
overflows and transfers a large amount of tokens on the attacker's behalf. 
This vulnerability can be formalized as follows:
\begin{vul}{{\bf BatchOverflow}}  
\begin{equation}
\begin{split}
\exists arg_0, arg_1, r_1, r_2, r_3, call \\
           r_3 = (r_1 \otimes r_2) \ \land \ \geval{r_2} > \geval{r_3} \land
           (\text{interfere?} \ r_2 \ \text{call.value}) \ \land \\
           (\text{interfere?} \ arg_0 \ \text{call.addr}) \ \land  
           (\text{interfere?} \ arg_1 \ \text{call.value})
\end{split}
\end{equation}
\end{vul}
\noindent In other words, the victim program contains a \texttt{call} instruction whose beneficiary and value 
can be  controlled by the attacker. Furthermore, the transaction value is also influenced by a variable 
from an arithmetic operation that overflows.

A Timestamp Dependency vulnerability occurs if a transaction depends on a timestamp: 
\begin{vul}{{\bf Timestamp Dependency}} 
\begin{equation}
\begin{split}
\exists \ \text{timestamp}, \text{call}. \ \text{call.value} > 0 \ \land \\
          (\text{interfere?} \ \text{timestamp} \ \text{call.value})
\end{split}
\end{equation}
\end{vul}
\noindent This vulnerability enables a malicious miner to gain an advantage by choosing a suitable timestamp for a block.

For some critical instructions such as \texttt{delegatecall} and \texttt{call}, runtime errors
will not lead to a rollback of the current state and the programmer is responsible for manually 
checking the return values and restoring the program state. Failing to do so can lead to an \emph{Unchecked-send Vulnerability} with unexpected
behavior~\cite{gasless}. We formalize the absence of this check as follows: 
\begin{vul}{{\bf Unchecked-send (Gasless-send)}}
\begin{equation}
\begin{split}
\neg \forall \ \text{call.ret}, \exists \ \text{jmp.var} 
   (\text{interfere?} \ \text{call.ret} \ \text{jmp.var})
\end{split}
\end{equation}
\end{vul}
\noindent Here, the return value of a \texttt{call} instruction does not 
\emph{interfere with} the conditional variables of any \emph{conditional jump} statements. 
In other words, this return value is not checked.

In section~\ref{sec:intro}, we briefly introduce the Reentrancy vulnerability. 
This vulnerability occurs when an attacker's call is allowed to 
repeatedly make new calls to the same victim contract without updating the victim's balance. 
It can be overapproximated as follows:
\begin{vul}{{\bf Reentrancy}} 
\begin{equation}
\begin{split}
\exists \ \text{arg}, i, j, k, l. \ i + 1 = j \ \land \ j < k \\
           l[i] = \text{``call"} \land l[j] = \text{``call"} \ \land l[k] = \text{``store"} \ \land \\
           l[i].\text{gas} > 2300 \land (\text{interfere? arg} \ l[i].\text{addr})\\
           \text{where } l \text{ is an execution trace.}
\end{split}
\end{equation}
\end{vul}
\noindent In other words, if an attack program has the minimum gas (i.e., 2300) to 
control the recipient of a transaction and generate consecutive \texttt{call} instructions before updating the storage, there may exist a Reentrancy vulnerability.
\subsection{Attack Synthesis.}\label{sec:attack}

Given a vulnerability query, we are interested in synthesizing an attack 
program that can exploit this vulnerability in a victim contract. 
The basic building blocks of an attack program are called 
\emph{components}, and each component $\comp$ corresponds to a public function provided by a 
victim contract. We use $\comps$ to denote the union of all publicly available 
methods.

\begin{definition}{{(\bf Component)}}
A \emph{Component} $\comp$ from an ABI configuration is a pair $(f, \tau)$ where:
\begin{itemize}
\item $f$ is $\comp$'s name. 
\item $\tau$ is the type signature of $\comp$.
\end{itemize}
\end{definition}

\begin{example}
Considering the ABI configuration in Figure~\ref{fig:attack-abi}, its first element 
(line 2-12) declares a component for the problematic \texttt{batchTransfer} method in 
figure~\ref{fig:attack-vic}. In particular, this component takes inputs as an 
array of \texttt{address} and a 256-bit integer (\texttt{uint256}). 
\end{example}


We represent a set of candidate attack programs as a \emph{symbolic program}, 
which is a sequence of \emph{holes} to be filled with components from $\comps$. 
The synthesizer fills these holes to obtain 
a \emph{concrete program} that exploits a given vulnerability.
\begin{definition}{{(\bf Symbolic Attack Program)}}\label{def:symbolic-attack}
Given a set of components $\comps=\{(f_1,\tau_1),\ldots,(f_N,\tau_N)\}$, a \emph{symbolic attack program} $\sketch$ 
for $\comps$ is a sequence of \emph{statement holes} of the form\looseness=-1
$$
\mathtt{choose}(f_1({\vec{v}_{\tau_1}}), \ldots, f_N({\vec{v}_{\tau_N}}));
$$
where $f_i({\vec{v}_{\tau_i}})$ stands for the application of the
$i$-th component to fresh symbolic values of types specified by $\tau_i$. 
\end{definition}
\begin{definition}{{(\bf Concrete Attack Program)}}
A \emph{concrete attack program} for a symbolic program $\sketch$ 
replaces each hole in $\sketch$ with one of the specified function calls, 
and each symbolic argument to a function call is replaced with a concrete value.
\end{definition}
\begin{example}
Here is a symbolic program that captures the attack candidate in
Fig~\ref{fig:attack-code}:
\begin{lstlisting}[numbers=none,frame=none,basicstyle=\footnotesize\ttfamily]
choose(makeFlag($x_1$), batchTransfer($y_1$,$z_1$));  
choose(makeFlag($x_2$), batchTransfer($y_2$,$z_2$)); 
\end{lstlisting}
And here is a concrete attack program for this symbolic attack:  
\begin{lstlisting}[numbers=none,frame=none,basicstyle=\footnotesize\ttfamily]
makeFlag(true); 
batchTransfer([0x123,0x345], $2^{256}-1$);  
\end{lstlisting}

\end{example}

Note that we use the $\mathtt{choose}$ construct to represent holes in symbolic
programs only for notational convenience. Since our smart contract language
supports symbolic values, every instance of $\mathtt{choose}$ can be expressed
using a conditional statement that guards the specified choices with fresh
symbolic booleans. For example, $\mathtt{choose}(e_1, e_2)$ is a notational
shorthand for the statement $\mathtt{if}\ b_1\ \mathtt{then}\ e_1\
\mathtt{else}\ e_2$, where $b_1$ is a fresh symbolic boolean value. A concrete
attack program therefore substitutes concrete values for the implicit $\mathtt{choose}$ guards and the
explicit function arguments of a symbolic attack program. So, 
all attack programs are expressible in our smart contract language with no extra
machinery. 




Since attack programs are valid programs in our language, 
their semantics is given by the rules in Figure~\ref{fig:sem}. 
We write $\peval{\sketch_0;...\sketch_n;}$ to represent the result of executing 
the attack program $\sketch = \sketch_0;...\sketch_n;$ from the program state $\pstate$.
If $\sketch$ is a symbolic attack program, then $\peval{\sketch_0;...\sketch_n;}$
represents the states $\pstate^{*}$ reachable by all concrete programs for $\sketch$ starting from the state $\pstate$. 
The goal of attack synthesis is to  
find a concrete program $P$ for a given symbolic program $\sketch$ 
such that $P$ reaches a state satisfying a desired vulnerability query.

\begin{definition}{{(\bf Problem Specification)}}
The specification for our \emph{attack synthesis} problem is a tuple ($\pstate_0$, $\vulnerability$, $\sketch$) where:
\begin{itemize}
\item $\sketch$ is a symbolic attack program for the set of components $\comps$
of a victim contract $\victim$.
\item $\pstate_0$ is the initial state of the symbolic attack program, obtained by executing the victim's initialization code.
\item $\vulnerability$ is a first-order formula over program states
$\pstate^{*}$ reachable from $\pstate_0$ by the attack program $\sketch$.
\end{itemize}
\end{definition}
\begin{definition}{{(\bf Attack Synthesis)}}
Given a specification ($\pstate_0$, $\vulnerability$, $\sketch$), the 
\emph{Attack Synthesis problem} is to find a \emph{concrete attack program} $P$ for
$\sketch$ such that:
\begin{itemize}
\item $\geval{ P }_{\pstate_0} = \pstate$
\item $\pstate \models \vulnerability$
\end{itemize}
In other words, executing 
the concrete attack $P$ from the initial state $\pstate_0$
results in a program state $\pstate$ that satisfies $\vulnerability$.
\end{definition}

\section{Summary-based Symbolic Evaluation}\label{sec:sum}

Solving the attack synthesis problem involves searching for a concrete program $P$  
in the space of candidate attacks defined by a symbolic program $\sketch$. 
\toolname delegates this search to an off-the-shelf SMT solver, 
by using symbolic evaluation to reduce the attack synthesis problem to a satisfiability query. 
Given a specification  $(\pstate_0, \vulnerability, \sketch)$,  \toolname evaluates $\sketch$ on the state $\pstate_0$ to 
obtain the state $\geval{\sketch}_{\pstate_0}$, and then uses the solver to check the 
satisfiability of the  formula $\exists \vec{v} . \vulnerability(\geval{\sketch}_{\pstate_0})$, where 
$\vec{v}$ denotes the symbolic variables in $\sketch$. 
A model of this formula, if it exists, binds every variable in $\vec{v}$ to a concrete value, 
and so represents a concrete attack program $P$ for $\sketch$ that triggers the vulnerability $\vulnerability$.

But computing $\geval{\sketch}_{\pstate_0}$ is expensive, as it relies on symbolic execution~\cite{rosette}. 
In particular, evaluating  $\sketch$ involves evaluating each of its $\mathtt{choose}$ statements, 
which, in turn, requires symbolically executing each function call in that statement. 
So, for a symbolic program of length $K$, every public function in the victim contract 
must be symbolically executed $K$ times on different symbolic arguments. 
As we will see in section~\ref{sec:eval}, this direct approach to evaluating $\sketch$ does not scale to 
real contracts that contain a large number of complex public functions.  
To mitigate this issue, we use a summary-based 
symbolic evaluation that performs symbolic execution of each public method only once. 

Our approach is based on the following insight. 
An attack program performs a sequence of transactions---i.e., method invocations---that 
manipulate the victim's persistent storage and global properties. 
The transactions that comprise an attack exchange data and influence each other's control flow 
exclusively through these two parts of the program state. 
So, if we can faithfully summarize the effects of a public method on 
the persistent storage and global properties,
evaluating this summary on the symbolic arguments passed to the method  
is equivalent to symbolically executing the method itself.\looseness=-1

\begin{definition}
A summary $\summary$ in our system is a pair $s@\pc$ where $s$ represents a statement that
has a side effect on the persistent state (i.e., storage and global properties) of a smart contract, and $\pc$ 
denotes the path condition of executing $s$.
\end{definition}

We generate such faithful method summaries in two steps. First, we use the rules
in Figure~\ref{fig:grammar} to execute the method on a program state $\pstate_A$
that maps every state variable (i.e., persistent storage location, global
property, etc.) to a fresh symbolic variable of the right type. This symbolic
execution step produces a path condition and symbolic inputs for each
instruction that capture every possible way to reach and execute the instruction
within the given method.
Next, we use the procedure in Figure~\ref{fig:sum-gen} to generate the method
summary.\footnote{We omitted the details of other instructions such as
\texttt{selfdestruct} and \texttt{delegatecall}.} Given a storage-store
instruction \texttt{sstore(x,y)} and its path condition, we generate a ``summary
sstore" statement (i.e., $\sumi{sstore}$) that takes as input the name of the
storage variable (i.e., $x$) and the symbolic expression $\eval{y}$ held in the
register $y$. Similarly, given a \texttt{call(gas,addr,value)} instruction and
path condition, we emit its ``summary call" statement (i.e., $\sumi{call}$) that
takes as input the symbolic expressions of the instruction's gas consumption,
recipient address, and amount of cryptocurrency, respectively. All other
instructions are omitted from the summary since they have no effect on the
persistent state. By construction, our summary therefore precisely  captures all
of the method's effects on the persistent state.\looseness=-1

\begin{example}
Figure~\ref{fig:sum-interp}c shows the summary of the program in Figure~\ref{fig:sum-interp}b.
Using the rule in Figure~\ref{fig:sum-gen}, our tool summarizes the side effects of the \texttt{call} and \texttt{sstore} instructions 
at lines 3 and 4, respectively.  The remaining instructions (lines 6--9) are omitted from the summary.\looseness=-1
\end{example}

Given a method summary and a program state $\pstate$, we use the procedure in
Figure~\ref{fig:sum-inter} to reproduce the effects of executing the method
symbolically on $\pstate$ as follows. Recall that we generate the summary by
executing the method on a fully symbolic state $\pstate_A=\{x_1\mapsto
v_1,\ldots, x_n\mapsto v_n\}$, so every path
condition and symbolic expression in the summary is given in terms of the
symbolic variables $v_1,\ldots,v_n$. Our summary interpretation procedure works by
substituting each $v_i$ in an instruction's path condition and inputs with its
corresponding value in $\pstate$, i.e., $\pstate[x_i]$. The resulting instruction summary
$s_{\pstate}@\pc_{\pstate}$ is therefore expressed in terms of $\pstate$, so applying its
side effects $s_{\pstate}$ under the path condition $\pc_{\pstate}$ is equivalent to
executing the instruction $s$ in the original method on the state $\pstate$. 
Since we interpret every instruction in the summary in this way, the
combined effect on the persistent state is equivalent to executing the original
method symbolically on $\pstate$.\looseness=-1


\begin{example}
Figure~\ref{fig:sum-interp}d shows an example for interpreting the summary in
Figure~\ref{fig:sum-interp}c by applying the procedure in
Figure~\ref{fig:sum-inter}. Specifically, given the \texttt{call} summary at
line 2 in Figure~\ref{fig:sum-interp}c, we first generate an \texttt{if}
statement guarded by the path condition $\pc$ for the current
summary, then in the body of the \texttt{if} statement we symbolically simulate
the side effect of the \texttt{call} statement by applying the \texttt{call}
rule in Figure~\ref{fig:sem}.
\end{example}


\begin{figure}
    \begin{subfigure}[b]{0.4\textwidth}
		\begin{lstlisting}
(define (get-summary s $\pc$)
  (match s
   [call(x, y, z) $\sumi{call}$($\eval{x}$, $\eval{y}$, $\eval{z}$)@$\pc$]
   [sstore(x, y)  $\sumi{sstore}$(x, $\eval{y}$)@$\pc$]
   [_             $\eval{\_}$)]))
		\end{lstlisting}
		\caption{Procedure for Summary Generation}
      \label{fig:sum-gen}
    \end{subfigure} \\
    \begin{subfigure}[b]{0.4\textwidth}
    \begin{lstlisting}      
(define (interpret-summary $s$@$\pc$ ${\pstate}$) 
  (define $s_{\pstate}$@$\pc_{\pstate}$ (substitute $s$@$\pc$ ${\pstate}$))
  (match $s_{\pstate}$
   [$\sumi{call}(x_{\pstate}, y_{\pstate}, z_{\pstate})$  (when $\pc_{\pstate}$ call($x_{\pstate}$, $y_{\pstate}$, $z_{\pstate}$))]
   [$\sumi{sstore}(x_{\pstate}, y_{\pstate})$     (when $\pc_{\pstate}$ sstore($x_{\pstate}$, $y_{\pstate}$))]
   [_  no-op]))
\end{lstlisting}
		\caption{Procedure for Summary Interpretation}
      \label{fig:sum-inter}
    \end{subfigure}
    		\caption{Summary Generation \& Interpretation}
      \label{fig:sum-ciai}
\end{figure}
\section{Implementation}\label{sec:impl}

This section discusses the design and implementation of 
\toolname, as well as two key optimizations that enable our tool 
to efficiently solve the synthesis attack problem.\looseness=-1

\subsection{Symbolic Computation Using \rosette}\label{sec:rosette}

\toolname leverages \rosette~\cite{rosette} to symbolically search for attack
programs. \rosette is a programming language that provides facilities for
symbolic evaluation.  These facilities are based on three constructs:
assertions, symbolic values, and (satisfiability) queries. \rosette programs use
assertions and symbolic values to formulate queries about program behavior,
which are then solved with off-the-shelf SMT solvers. For example, the
\texttt{(solve expr)} query searches for a binding of symbolic variables to
concrete values that satisfies the assertions encountered during the symbolic
evaluation of the program expression \texttt{expr}. \toolname uses the \texttt{solve}
query to search for a concrete attack program.

\begin{figure}
\begin{lstlisting}      
(define (smartscopy $\vulnerability$  $\comps$ $K$)
 (define (stmt) (apply choose* $\comps$))
 ;;Generate a symbolic attack program of size $K$.
 (define program (map ($\lambda$ (x) (stmt)) (range $K$)))
 (define (progstate)
  ;;Program state has registers, memory, storage, 
  ;;gas, and other global information.
  (progstate (for/vector ([i config]) 'reg)
             (init-memory)
             (init-storage)
             'gas  ;;gas consumption
             ...))
 (define i-pstate (send machine get-state ...))
 (define o-pstate (interpret program i-state))
 (define binding (solve (assert ($\vulnerability$ o-pstate))))
 (evaluate program binding))
\end{lstlisting}
\caption{\toolname implementation in \rosette. 
 }
\label{fig:sketch-overview}
\end{figure}

Figure~\ref{fig:sketch-overview} shows the implementation of \toolname in Rosette. 
The tool takes as input a vulnerability specification
$\vulnerability$, the components $\comps$ of a victim program, and a bound $K$ 
on the length of the attack program. Given these inputs, lines 2--4 use $\comps$ to 
construct a symbolic attack \texttt{program} of length $K$. 
Next, lines 5--13 run the victim's initialization code to obtain the 
initial program state, \texttt{i-pstate}, for the attack. 
Then, line 14 evaluates the symbolic attack \texttt{program} on the initial state to 
obtain a symbolic output state, \texttt{o-pstate}. 
Finally, lines 15-16 use the \texttt{solve} query to search for a concrete
attack program that satisfies the vulnerability assertion.

The core of our tool is the \emph{interpreter} for our smart contract language
(Figure~\ref{fig:grammar}), which implements the operational semantics given in
Figure~\ref{fig:sem}. We use this interpreter to compute the symbolic summaries of the
victim's public methods (Section~\ref{sec:sum}) and to evaluate symbolic attack programs. 
The interpreter itself does not implement symbolic execution; instead, it uses 
\rosette's symbolic evaluation engine to execute programs in our language on
symbolic values. 

Another key component of \toolname is the \emph{translator} that converts EVM
bytecode into our language (Figure~\ref{fig:grammar}). The translator leverages the Vandal
Decompiler~\cite{madmax} to soundly convert the stack-based EVM bytecode into
its corresponding three-address format in our language. The jump targets are resolved through abstract
interpretation~\cite{CousotC77}.  We use the translator to convert victim
contracts to the \toolname language for attack synthesis. Both the translator and the
interpreter support all the instructions defined in the Ethereum
specification~\cite{yellowpaper}.\looseness=-1


\subsection{Parallel Synthesis using Hoisting}\label{sec:parallel}

\toolname uses summary-based symbolic evaluation to efficiently reduce attack
synthesis problems to satisfiability queries. But the resulting queries can
still be too difficult (for both \rosette and the underlying solver) to solve in
practice, especially when the victim contract has many public methods.
So to further improve performance, \toolname exploits the structure of symbolic
attack programs (Definition~\ref{def:symbolic-attack}) to decompose the single 
\texttt{solve} query in Figure~\ref{fig:sketch-overview} into multiple smaller 
queries that can be solved quickly and in parallel, without missing any concrete attacks.

The basic idea is as follows. Given a set of $N$ components and a bound $K$ on 
the length of the attack, lines 2--4 create a symbolic attack program of the
following form:
\begin{lstlisting}[numbers=none,frame=none,basicstyle=\footnotesize\ttfamily]
  choose$_1$($f_1({\vec{v_1}_{\tau_1}}), \ldots, f_N({\vec{v_1}_{\tau_N}})$);  
  $\vdots$
  choose$_K$($f_1({\vec{v_k}_{\tau_1}}), \ldots, f_N({\vec{v_K}_{\tau_N}})$);   
\end{lstlisting}
This symbolic attack encodes a set of concrete attacks that can also be
expressed using $N^K$ symbolic programs that fix the choice of the method to
call at each line, but leave the arguments symbolic. So, we can enumerate these
$N^K$ programs and solve the vulnerability query for each of them, instead of
solving the single query at line 15. This approach essentially \emph{hoists} the
symbolic boolean guards out of the \texttt{choose} statements in the original query, and
\toolname explores all possible values for these guards explicitly, rather than
via SMT solving. 
As we show in Section~\ref{sec:eval},
hoisting the guards leads to significantly faster synthesis, 
both because it enables parallel solving of the smaller queries, and 
because the smaller queries can be solved quickly.\looseness=-1

\subsection{SMT-based Early Pruning}\label{sec:prune}

In addition to hoisting, we also design a simple but effective \emph{early pruning}
strategy that allows \toolname to prune \emph{infeasible} symbolic programs 
before executing them. The intuition behind our early pruning strategy is that
all attacks expressible in \toolname (e.g.,~\cite{attack1,attack2,attack3}) invoke at least one public method that 
manipulates persistent storage and at least one public method that transfers 
cryptocurrency using the \texttt{call} instruction. In other words,  
a successful attack executes at least one store instruction followed 
by at least one \texttt{call} instruction. We express our early pruning strategy using the following 
\rosette program:
\begin{lstlisting}
(define (may-store-and-call? p)
 (solve (exists (list i j)
  (and (< i j) (= (type p[i]) 'store)
	       (= (type p[j]) 'call)))))
\end{lstlisting}
This procedure queries the solver to find out if the given symbolic program $p$ contains 
any concrete attack program that executes a \texttt{call} after a store. 
This query is much faster to solve than a vulnerability query, so if $p$ contains no 
feasible candidate, \toolname does not run the vulnerability query for it.\looseness=-1

\section{Evaluation}\label{sec:eval}
We evaluated \toolname by conducting two experiments that are designed to answer the following questions: 
\begin{itemize}
\item Q1: \emph{Expressiveness}: Can \toolname express the specifications of real world vulnerabilities? 
\item Q2: \emph{Effectiveness}: How does \toolname compare against state-of-the-art analyzers for smart contracts?
\item Q3: \emph{Efficiency}: How much does summary-based symbolic 
evaluation improve the performance of \toolname?
\end{itemize}

To answer these questions, we perform a systematic evaluation by running \toolname on the entire set of smart contracts from Etherscan~\cite{etherscan}. Using a snapshot from August 30 2018, we obtained a total number of 25,983 smart contracts. Similar to the teEther~\cite{teether}
paper, we restrict the maximum size of our attack programs to three.
All experiments in this section are conducted on a \texttt{t3.2xlarge} machine on Amazon EC2 with an Intel Xeon Platinum 8000 CPU and 32G of memory, running the Ubuntu 18.04 operating system and using a timeout of 10 minutes for each smart contract.

\subsection{Expressiveness of \toolname}\label{sec:express}
To understand the expressiveness of our tool, we encoded the common vulnerabilities in 
smart contracts described in prior work~\cite{attack-time,smart-sec} and on social 
media~\cite{batch-news}. 
In particular, Table~\ref{tbl:expr} summarizes the expressiveness of existing tools
for Smart Contract security, ordered by publish date. Note that our tool  
 supports not only well-known vulnerabilities such as Reentrancy, Timestamp Dependency,
 and Arithmetic operations (i.e., over/underflow), but also recent attacks
 such as the short address attack and the BatchOverflow vulnerability discussed 
 in Section~\ref{sec:overview}. Prior tools express 
 a portion of these vulnerabilities. For instance, the popular \oyente~\cite{oyente} 
 tool, which is also based on symbolic execution, does not support vulnerabilities such as 
 unchecked calls, short address, and out-of-gas-DoS. Static analysis tools such as 
 Securify~\cite{securify} and \madmax~\cite{madmax} do not support complex arithmetic vulnerabilities. Most importantly, unlike \toolname, none of them can generate exploits for vulnerabilities. The teEther and the \contractfuzz tools can automatically generate 
 exploits, but their systems only support a small class of vulnerabilities.

 There are some vulnerabilities that our tool does not support well. 
 For instance, a Transaction-Ordering Dependency (TOD) is a race condition 
 vulnerability, and exploiting it requires synthesizing \emph{a pair of programs}
 that exhibit the race. In the future, we plan to explore \emph{relational synthesis}
 to handle attacks that require multiple programs. Another source of limitation
 is denial-of-service (DoS) attacks that involve loops, which our tool unrolls during 
 symbolic execution, and the unrolling bound may not be big enough to
 trigger the vulnerability.

\begin{table*}[]
\centering
\footnotesize
\begin{tabular}{|l|c|c|c|c|c|c|c|c|c|c|}
\hline
\multicolumn{1}{|c|}{\multirow{2}{*}{Tool}} & \multicolumn{1}{l|}{\multirow{2}{*}{\begin{tabular}[c]{@{}l@{}}Generate\\ Exploit?\end{tabular}}} & \multicolumn{9}{c|}{Common Vulnerabilities}                                                                                                                                                                                                                                                                                                                                                                                                                                                                                                                              \\ \cline{3-11} 
\multicolumn{1}{|c|}{}                      & \multicolumn{1}{l|}{}                                                                             & \multicolumn{1}{l|}{Reentrance} & \multicolumn{1}{l|}{Arithmetic} & \multicolumn{1}{l|}{DoS} & \multicolumn{1}{l|}{\begin{tabular}[c]{@{}l@{}}Bad \\ Random\end{tabular}} & \multicolumn{1}{l|}{Timestamp} & \multicolumn{1}{l|}{TOD} & \multicolumn{1}{l|}{\begin{tabular}[c]{@{}l@{}}Unchecked \\ Calls\end{tabular}} & \multicolumn{1}{l|}{\begin{tabular}[c]{@{}l@{}}Access \\ Control\end{tabular}} & \multicolumn{1}{l|}{\begin{tabular}[c]{@{}l@{}}Short \\ Address\end{tabular}} \\ \hline
\oyente~\cite{oyente}                       & \LEFTcircle                                                                                    & \cmark                          &\cmark                            &                    &                                                                      & \cmark                          & \LEFTcircle                   &                                                                           &                                                                          &                                                                                                                                                \\ \hline
Mythril~\cite{mythril}                      &  \LEFTcircle                                                                                           & \cmark                           & \cmark                          &                    &                                                                      & \cmark                          & \LEFTcircle                  & \cmark                                                                          & \cmark                                                                         &                                                                                                                                                 \\ \hline

Zeus~\cite{zeus}                            &                                                                                             & \cmark                          & \cmark                          &                    &                                                                      & \cmark                          & \LEFTcircle                   & \cmark                                                                          & \cmark                                                                         &                                                                                                                                                 \\ \hline

teEther~\cite{teether}                      & \cmark                                                                                            &                           &                           &                    &                                                                      &                          &                    &                                                                           &                                                                          &                                                                                                                                                 \\ \hline

Securify~\cite{securify}                    &                                                                                             & \cmark                          &                           &                    &                                                                      & \cmark                          & \LEFTcircle                   & \cmark                                                                          & \cmark                                                                         &                                                                                                                                                  \\ \hline
\madmax~\cite{madmax}                       &                                                                                             &                           & \LEFTcircle                           & \LEFTcircle                    &                                                                      &                          &                    &                                                                           &                                                                          &                                                                                                                                                  \\ \hline
ContractFuzzer~\cite{contractfuzzer}        & \cmark                                                                                            & \cmark                          &                           &                    &\cmark                                                                      & \cmark                          & \LEFTcircle                  &                                                                           &                                                                          &                                                                                                                                                 \\ \hline

\toolname                                   & \cmark                                                                                            & \cmark                          & \cmark               & \LEFTcircle                   & \cmark                                                                     & \cmark                         & \LEFTcircle                   & \cmark                                                                          & \cmark                                                                         & \cmark                                                                                                                                                \\ \hline

\end{tabular}
\caption{A Comparison of Existing Tools for Smart Contract (Order by publish date). \LEFTcircle \ represents limited support.}. 
\label{tbl:expr}
\end{table*}

\subsection{Comparison with Existing Tools}\label{sec:comp}
To demonstrate the advantages of our proposed approach, we compare \toolname against two
state-of-the-art analyzers that are publicly available: \oyente, based on 
symbolic execution, and 
\contractfuzz, based on dynamic random testing~\footnote{Other tools like teEther and Securify are not available for comparison at the time of this submission.}.

\paragraph{Comparison with \oyente}\label{sec:oyente}


\begin{table}[]
\centering
\begin{tabular}{|l|l|l|l|}
\hline
\multicolumn{1}{|c|}{\multirow{2}{*}{Vulnerability}} & \multicolumn{3}{l|}{Number of vulnerable contracts} \\ \cline{2-4} 
\multicolumn{1}{|c|}{}                               & \multicolumn{1}{c|}{$S \land O$}            & \multicolumn{1}{c|}{$S - O$}           & \multicolumn{1}{c|}{$O - S$}           \\ \hline
Timestamp   &\multicolumn{1}{c|}{485}  &\multicolumn{1}{c|}{842}   &\multicolumn{1}{c|}{39}                \\ \hline
Reentracy   &\multicolumn{1}{c|}{49}  &\multicolumn{1}{c|}{128}   &\multicolumn{1}{c|}{41}                \\ \hline
\end{tabular}
\caption{Comparing \toolname ($S$) against \oyente ($O$).
$S \land O$, $S - O$, and $O - S$ represent \# of benchmarks 
reported by both tools, $S$ only, and $O$ only, respectively.}
\label{fig:eval-oyente}
\vspace{-0.1in}
\end{table}

We first compare with \oyente~\cite{oyente}, which takes as input 
a smart contract and checks whether there are concrete traces that match
the tool's predefined security properties. If so, the tool returns a counterexample
as the exploit. We evaluate \oyente and \toolname on the Etherscan data set, and 
both systems use a timeout of ten minutes.

The \oyente tool supports four different types of vulnerabilities, namely, 
call-stack-limit, Timestamp dependency~\cite{attack4}, Reentrancy~\cite{attack1}, 
and Transaction-Ordering dependency (TOD)~\cite{attack4}. Since the call-stack-limit
vulnerability had already been fixed by the Solidity team and the TOD vulnerability
requires synthesizing multiple programs, we will cover the 
remaining two vulnerabilities.

\paragraph{Summary of results}
The results of our evaluation are summarized in Table~\ref{fig:eval-oyente}. In particular, for the Timestamp dependency vulnerability, there
are 485 benchmarks where both tools report a vulnerability and find the exploits.
39 benchmarks are flagged as vulnerable by \oyente but \toolname can not find 
the exploits. We manually inspected the source code of those benchmarks and 
confirm that 30 of them are false positives. On the other hand, 842
benchmarks are flagged as safe by \oyente while \toolname manages to find
their exploits. To verify the reports of our tool, we randomly select 
20 benmarks and confirm 18 of them are actually vulnerable. In the meantime,
we also contacted the author of \oyente and confirmed our report.

For the Reentrancy vulnerability, 49 benchmarks are flagged by both tools.
41 benchmarks are flagged as vulnerable by \oyente while \toolname cannot
find the exploits. After manual inspection, we confirm all of them are false 
positives. In contrast, 128 benchmarks are marked as safe but \toolname 
successfully finds their exploits, and we manage to reproduce 102 of the attacks 
in our testbed.
\begin{table}[]
\centering
\begin{tabular}{|c|c|c|c|c|}
\hline
\multirow{2}{*}{Vulnerability} & \multicolumn{2}{c|}{\toolname} & \multicolumn{2}{c|}{\oyente} \\ \cline{2-5} 
                               & FP             & FN            & FP            & FN           \\ \hline
Timestamp                      & 7\%            & 10\%            & 36\%             & 35\%            \\ \hline
Reentrancy                     & 14\%             & 5\%           & 43\%             & 37\%            \\ \hline
\end{tabular}
\caption{Analysis of the results based on full inspection on 20 random samples 
from $S \cup O$}
\label{fig:eval-oyente-fp-fn}
\end{table}

To further understand the effectiveness of both tools, we randomly pick 20 
samples from a subset of the data where each contract is flagged as vulnerable
by at least one tool. We repeat this process three times and report the average.
As shown in Table~\ref{fig:eval-oyente-fp-fn}, for 
the Timestamp vulnerability, the FN and FP rates of \toolname are 7\% and 10\%,
while the FN and FP rates of \oyente on our selected data set are 36\% and 35\%.
The result on the Reentrancy vulnerability is similar: the FN and FP rates of \toolname are 14\% and 5\%,
while the FN and FP rates of \oyente  are 43\% and 37\%.

\paragraph{Performance}
\oyente takes an average of 91 seconds to analyze a contract, while \toolname only takes an average of 8 seconds for this data set.
\paragraph{Discussion} 
To understand why \oyente has higher false positive and negative rates than
\toolname, we manually inspected 20 randomly chosen samples from each category. 
The results of this analysis are as follows.

The high false negative rate in \oyente is caused by low coverage on the
corresponding benchmarks. Specifically, in the presence of large and complex
benchmarks, \oyente fails to generate traces that trigger the vulnerability.
Moreover, since the Keccak-256 hash function is ubiquitous in smart contracts,
and hard for the solver to reason about, \oyente fails to cover the code regions
that have dependencies on the hash function. 

The false positives in \oyente can be attributed to two root causes. The first
is that the tool does not model the semantics of the gas system, and its query
language cannot reason about gas consumption in a smart contract. For instance,
\oyente will report spurious Reentrancy vulnerabilities even though the gas
specified by the victim is insufficient for an attacker to generate the exploit.
On the other hand, since \toolname precisely models the semantics of the gas
system, we are able to achieve a low false positive rate. The second cause of
false positives is due to the exploration of paths that an attacker cannot
trigger. 
For instance, \oyente marks the following code as Reentrancy vulnerability 
even though an attacker has no permission to trigger it.
\begin{lstlisting}[escapechar=@]
public function mintETHRewards(
  address _contract, uint256 _amount) 
  @\textbf{onlyManager}@() {
  require(_contract.call.value(_amount)());}
\end{lstlisting}

We also investigated the cause of false positives 
reported by \toolname. It turns out that the false positives are 
caused by the imprecision of our queries. Recall from Section~\ref{sec:vul}
that we use a specific pattern of traces to \emph{overapproximate} the
behavior of the Reentrancy attack. While effective and 
efficient in practice, our query may generate spurious 
exploits that are infeasible. To mitigate this limitation, one 
compelling approach for developing secure smart contracts is to 
ask the developers to provide invariants for preventing the vulnerabilities, 
and then use \toolname to search for exploits that violate the invariants. 

\subsection{Comparison with \contractfuzz}\label{sec:fuzz}
We further compared \toolname against \contractfuzz~\cite{contractfuzzer}, a recent 
smart contract analyzer based on dynamic fuzzing. Specifically, 
\contractfuzz takes input as the ABI interfaces of smart
contracts and \emph{randomly} generates inputs invoking the public methods 
provided by the ABI. To verify the correctness of the exploits, \contractfuzz
implements oracles for different vulnerabilities by instrumenting
the Ethereum Virtual Machine (EVM) with extra assertions.


\begin{table}[]
\centering
\begin{tabular}{|l|l|l|l|l|l|l|}
\hline
\multirow{2}{*}{Vulnerability} & \multicolumn{3}{l|}{\toolname} & \multicolumn{3}{l|}{\contractfuzz} \\ \cline{2-7} 
                               & No.       & FP       & FN      & No.        & FP        & FN        \\ \hline
Timestamp                      & 16        & 0        & 1       & 13         & 4         &7         \\ \hline
Gasless Send                   & 17        & 0        & 0       & 14         & 3         & 6         \\ \hline
Bad Random                     & 9        & 0        & 0       & 5          & 1         & 5         \\ \hline
\end{tabular}
\caption{Comparing \toolname against \contractfuzz}
\label{fig:eval-fuzz}
\end{table}

We use the docker image~\cite{fuzz-docker} provided by the author of \contractfuzz.
The original paper does not discuss the performance of the tool, 
but from our experience, \contractfuzz is slow, taking more than 
10 mins to fuzz a smart contract. Since it would be time-consuming to 
run \contractfuzz on the Etherscan data set, we evaluate both tools 
on the 33 benchmarks from the \contractfuzz artifact~\cite{fuzz-data} plus another
67 random samples from Etherscan for which we know the ground truth.

\paragraph{Summary of results}
The results of our evaluation are summarized in Table~\ref{fig:eval-fuzz}.
In particular, for the timestamp dependency, \contractfuzz flags 13 benchmarks 
as vulnerable. However, 4 of them are false alarms, and it fails to detect 7 
vulnerable benchmarks. On the other hand, \toolname detects most of 
the benchmarks with only one false negative, which is caused by a timeout on the Vandal
decompiler~\cite{madmax}.\looseness=-1

Similarly, for the Gasless-send vulnerability, 14 benchmarks are flagged by \contractfuzz.
However, 3 of them are false positives, and 6 vulnerable benchmarks can not be detected 
within 10 minutes. In contrast, \toolname successfully generates exploits for 
all the vulnerable benchmarks.

\paragraph{Performance}
On average, it takes \contractfuzz 10 mins to analyze a smart contract. 
\toolname takes an average of 11 seconds on this data set.

\paragraph{Discussion}
The cause of false negatives in \contractfuzz is easy to understand as it is
based on random, rather than exhaustive, exploration of an extremely large
search space. So if there are relatively few inputs in this space that lead to
an attack, \contractfuzz is unlikely to find it within the given time bound (10
minutes). On the other hand, the false positives in \contractfuzz are caused by
the limited expressiveness of its assertion language. For instance, the Time
Dependency is defined as the following assertion in \contractfuzz: 
\[
\textbf{TimestampOp} \wedge (\textbf{SendCall} \vee \textbf{EtherTransfer})
\]
The assertion raises a Time Dependency vulnerability if the smart contract
contains the \texttt{timestamp} and \texttt{call} instructions. It is
easy to raise false alarms with this assertion if the \texttt{call} instruction 
does not depend on \texttt{timestamp}. On the other hand, the \texttt{interfere?} 
function enables \toolname to reason about this dependency precisely.

\subsection{Impact of Summary-based Symbolic Evaluation}\label{sec:expr}
\begin{table}[]
\centering
\begin{tabular}{|l|l|l|l|l|}
\hline
\multirow{2}{*}{$S^{\dagger}$-mean} & \multirow{2}{*}{$S^{\diamond}$-mean} & \multicolumn{3}{c|}{\# of Benchmarks Timeout} \\ \cline{3-5} 
                    &                      & $S^{\dagger} \land S^{\diamond}$   & $S^{\dagger} - S^{\diamond}$  & $S^{\diamond} - S^{\dagger}$ \\ \hline
8s                   & 35s                    & 1846         & 548        & 17454     \\ \hline
\end{tabular}
  \caption{Comparison between Summary-based ($S^{\dagger}$) and Non-summary ($S^{\diamond}$). $S^{\dagger} \land S^{\diamond}$, $S^{\dagger} - S^{\diamond},$ and $S^{\diamond} - S^{\dagger}$ represent number of benchmarks timeout on both, $S^{\dagger}$ only, and $S^{\diamond}$ only, respectively.}
  \label{fig:summary}
\end{table}

To understand the impact of our summary-based symbolic evaluation described 
in Section~\ref{sec:sum}, we run \toolname on the Etherscan data set with ($S^{\dagger}$) and 
without ($S^{\diamond}$) computing the summary. 
To speed up the evaluation, for both settings, we enable the early pruning and 
parallel synthesis optimizations discussed in Section~\ref{sec:impl}. 

As shown in Table~\ref{fig:summary},  if we exclude the benchmarks that timeout
in 10 mins, the mean time of our summary-based symbolic evaluation is only 8
seconds, while it takes 35 seconds without computing the summary. Furthermore,
1846 benchmarks time out for both settings, and only 548 benchmarks time out on
$S^{\dagger}$ but not on $S^{\diamond}$. However, without computing the summary,
17454 (i.e., 69.8\%) benchmarks time out. The result confirms that the
summary-based technique is key to the efficiency of \toolname.

\subsection{A case study on the recent BatchOverflow vulnerability}\label{sec:case}

To evaluate whether \toolname can discover new vulnerabilities in real world
smart contracts, we conduct a case study on the recent \emph{BatchOverflow
Vulnerability}. As we mentioned in Section~\ref{sec:overview}, exploits
due to this vulnerability have resulted in the creation of trillions of invalid
Ethereum Tokens in 2018~\cite{batch-news}, causing major exchanges to temporary
halt until all tokens could be reassessed. We note that generating exploits for
this vulnerability is quite challenging as it requires the tool to reason about
the combination of arithmetic operations, interference, and the read-write
semantics of the storage system in Solidity. For instance, existing tools such
as \oyente and \madmax~\cite{madmax} will simply mark a large number of arithmetic
operations as \emph{potentially vulnerable}, and it turns out that most of the
alarms are not exploitable.

Similar to our previous experiment, we first encode the vulnerability
(Section~\ref{sec:vul}) in our language and then run our tool on the 
Etherscan data set. \toolname generates exploits for 32 vulnerable
contracts. To verify that the exploits are effective, we setup a private
blockchain using the Geth~\cite{geth} framework where we can run exploits on the
vulnerable contracts. We confirmed that 20 exploits are valid. 
The infeasible attacks come from the incompleteness of the query 
as well as imprecise control flow graphs from the Vandal decompiler. Since
those contracts are covered by neither the previous literature nor the media, we
also sent the issues to their developers.
\section{Related Work}
Smart contract security has been extensively studied in recent years. 
In this section we briefly discuss prior closely related work.

\paragraph{Smart Contract Analysis}
Many popular security analyzers for smart contracts are based on symbolic execution~\cite{symbolic-e}.
Well-known tools include Oyente~\cite{oyente}, Mythril~\cite{mythril} and 
Manticore~\cite{manticore}. Their key idea is to find an execution path that satisfies
a given property or assertion. While \toolname also uses symbolic evaluation 
to search for attack programs, our system differs from these tools in two ways. First, 
the prior tools adopt symbolic execution for \emph{bug finding}. 
Our tool can be used not only for bug finding but also for \emph{exploit generation}. 
Second, while symbolic execution is 
a powerful and precise technique for finding security vulnerabilities, it does not 
guarantee to explore all possible paths, which leads to high false negative rates 
as shown in Section~\ref{sec:oyente}. In contrast, \toolname proposes a 
summary-based symbolic evaluation which significantly reduces the number of paths
it has to explore while maintaining the same precision.

To address the scalability and path explosion problems in symbolic execution, 
researchers developed sound and scalable static analyzers~\cite{ecf,securify,madmax,zeus}. 
Both Securify~\cite{securify} and Madmax~\cite{madmax} are based on abstract 
interpretation~\cite{CousotC77}, which soundly overapproximates and merges relevant 
execution paths to avoid path explosion. The ZEUS~\cite{zeus} system takes the source 
code of a smart contract and a policy as inputs, and then compiles them into LLVM IRs
that will be checked by an off-the-shelf verifier~\cite{smack}. The ECF~\cite{ecf} system
is designed to detect the DAO vulnerability. Similar to our tool, Securify also provides a query language to specify the patterns of common vulnerabilities. Unlike our tool, none of these systems can generate exploits. We could not directly compare \toolname with Securify and Zeus as their tools and benchmarks are not publicly available. However, we note that our system is complementary to existing static analyzers such as Securify: in particular, we can use Securify to filter out safe smart contracts and leverage \toolname to generate exploits 
for vulnerable ones.

Some systems~\cite{Hirai17,GrishchenkoMS18} for reasoning about smart contracts rely on formal verification. 
These systems prove security properties 
of smart contracts using existing interactive theorem provers~\cite{coq, isabelle}. 
They typically offer strong guarantees
that are crucial to smart contracts. However, unlike our system, all of them 
require significant manual effort to encode the security properties
and the semantics of smart contracts.

Finally, projects~\cite{porosity, madmax, Erays} related to reverse engineering aim to
to lift EVM bytecode to an intermediate representation that is easy to analyze.
Although \toolname uses the IRs from Vandal~\cite{madmax}, our technique is agnostic 
to the underlying language.

\paragraph{Automatic Exploitation}
Our work is also closely related to automatic exploitation~\cite{AvgerinosCHB11,ChaARB12,teether,contractfuzzer}. While those prior systems rely on constraint solvers to generate
counterexamples as potential exploits, we note that there are additional challenges 
in automatic exploitation for smart contracts. First, the exploits in classical vulnerabilities (e.g., buffer overflows, SQL injections) are typically program inputs of a specific data type (e.g., integer, string) whereas the exploits in our system are adversarial smart contracts that
faithfully model the execution environment (storage, gas, etc.) of the EVM. Second, Keccak-256 hash
is ubiquitous in smart contract for accessing addresses in memory or storage. As shown
in Section~\ref{sec:oyente}, basic symbolic execution will fail to resolve the Keccak-256 hash, 
resulting in poor coverage. To address this problem, the teEther~\cite{teether} system proposed a novel algorithm to infer the memory addresses encoded as Keccak-256 hash. Unlike
teEther, our system directly synthesizes function calls that manipulate the memory and storage 
thus avoids expensive computation to resolve the hash values.
We could not directly compare \toolname with teEther as its tools and benchmarks are not publicly available. 
Similar to \toolname, \contractfuzz~\cite{contractfuzzer} also generates exploits for 
a limited class of vulnerabilities based on the ABI specifications of smart contracts.
However, as shown in Section~\ref{sec:fuzz}, since \contractfuzz is completely
based on random input generation, it is an order of magnitude slower than \toolname.

\section{Conclusion}\label{sec:concl}
We presented \toolname, a tool for automatic synthesis of adversarial contracts 
that exploit the vulnerability of a victim smart contract under test. To make 
synthesis tractable, we introduced \emph{summary-based symbolic evaluation},
which significantly reduces the number of paths that our tool needs to 
explore while maintaining the precision of the vulnerability query. 
Building on the summary-based symbolic evaluation, \toolname further introduces 
optimizations that enable it to partition the synthesis search space for parallel exploration,
and to prune infeasible attack candidates earlier. We encoded common vulnerabilities of smart contracts in our query language, and 
evaluated \toolname on the entire data set from etherscan with $>$25K smart contracts. 
As shown in our experimental evaluation, \toolname significantly outperforms  state-of-the-art smart contract analyzers in terms of precision, soundness, and execution time.
Moreover, running on recent smart contracts, \toolname uncovers 20 previously 
unknown instances with the BatchOverflow vulnerability and none of the existing 
tool can precisely spot the vulnerability.

\bibliographystyle{unsrt}
\bibliography{main}

\begin{thebibliography}{10}

\bibitem{bitcoin}
Bitcoin.
\newblock \url{https://bitcoin.org/}, 2019.
\newblock [Online; accessed 01/09/2019].

\bibitem{ethereum}
Ethereum.
\newblock \url{https://www.ethereum.org/}, 2019.
\newblock [Online; accessed 01/09/2019].

\bibitem{solidity}
Solidity.
\newblock \url{https://solidity.readthedocs.io/en/v0.5.1/}, 2019.
\newblock [Online; accessed 01/09/2019].

\bibitem{etherscan}
Etherscan.
\newblock \url{https://etherscan.io/}, 2018.
\newblock [Online; accessed 01/09/2019].

\bibitem{case1}
Real estate business integrates smart contracts.
\newblock \url{https://tinyurl.com/yawrkfpx/}, 2018.
\newblock [Online; accessed 01/09/2019].

\bibitem{case2}
Smart contracts for shipping offer shortcut.
\newblock \url{https://tinyurl.com/yavel7xe/}, 2018.
\newblock [Online; accessed 01/09/2019].

\bibitem{attack1}
Understanding the dao attack.
\newblock \url{https://tinyurl.com/yc3o8ffk/}, 2017.
\newblock [Online; accessed 01/09/2019].

\bibitem{attack2}
On the parity wallet multisig hack.
\newblock \url{https://tinyurl.com/yca83zsg/}, 2017.
\newblock [Online; accessed 01/09/2019].

\bibitem{attack3}
Governmental’s 1100 eth payout is stuck because it uses too much gas.
\newblock \url{https://tinyurl.com/y83dn2yf/}, 2016.
\newblock [Online; accessed 01/09/2019].

\bibitem{attack4}
Nicola Atzei, Massimo Bartoletti, and Tiziana Cimoli.
\newblock A survey of attacks on ethereum smart contracts (sok).
\newblock In {\em Principles of Security and Trust - 6th International
  Conference, {POST} 2017, Held as Part of the European Joint Conferences on
  Theory and Practice of Software, {ETAPS} 2017, Uppsala, Sweden, April 22-29,
  2017, Proceedings}, pages 164--186, 2017.

\bibitem{oyente}
Loi Luu, Duc{-}Hiep Chu, Hrishi Olickel, Prateek Saxena, and Aquinas Hobor.
\newblock Making smart contracts smarter.
\newblock In {\em Proc. Conference on Computer and Communications Security},
  pages 254--269, 2016.

\bibitem{securify}
Petar Tsankov, Andrei~Marian Dan, Dana Drachsler{-}Cohen, Arthur Gervais,
  Florian B{\"{u}}nzli, and Martin~T. Vechev.
\newblock Securify: Practical security analysis of smart contracts.
\newblock In {\em Proc. Conference on Computer and Communications Security},
  pages 67--82, 2018.

\bibitem{contractfuzzer}
Bo~Jiang, Ye~Liu, and W.~K. Chan.
\newblock Contractfuzzer: fuzzing smart contracts for vulnerability detection.
\newblock In {\em Proc. International Conference on Automated Software
  Engineering}, pages 259--269, 2018.

\bibitem{madmax}
Neville Grech, Michael Kong, Anton Jurisevic, Lexi Brent, Bernhard Scholz, and
  Yannis Smaragdakis.
\newblock Madmax: surviving out-of-gas conditions in ethereum smart contracts.
\newblock In {\em Proc. International Conference on Object-Oriented
  Programming, Systems, Languages, and Applications}, pages 116:1--116:27,
  2018.

\bibitem{zeus}
Sukrit Kalra, Seep Goel, Mohan Dhawan, and Subodh Sharma.
\newblock {ZEUS:} analyzing safety of smart contracts.
\newblock In {\em Proc. The Network and Distributed System Security Symposium},
  2018.

\bibitem{teether}
Johannes Krupp and Christian Rossow.
\newblock teether: Gnawing at ethereum to automatically exploit smart
  contracts.
\newblock In {\em Proc. USENIX Security Symposium}, pages 1317--1333, 2018.

\bibitem{rosette}
Emina Torlak and Rastislav Bod{\'{\i}}k.
\newblock A lightweight symbolic virtual machine for solver-aided host
  languages.
\newblock In {\em Proc. Conference on Programming Language Design and
  Implementation}, pages 530--541, 2014.

\bibitem{attack-time}
Time manipulation.
\newblock \url{https://dasp.co/#item-8/}, 2018.
\newblock [Online; accessed 01/09/2019].

\bibitem{best-practice}
Ethereum smart contract security best practices.
\newblock \url{https://consensys.github.io/smart-contract-best-practices/},
  2018.
\newblock [Online; accessed 01/09/2019].

\bibitem{attack-int}
New batchoverflow bug in multiple erc20 smart contracts.
\newblock \url{https://tinyurl.com/yd78gpyt}, 2018.
\newblock [Online; accessed 01/09/2019].

\bibitem{yellowpaper}
Ethereum yellow paper.
\newblock \url{https://github.com/ethereum/yellowpaper}, 2019.
\newblock [Online; accessed 01/09/2019].

\bibitem{serpent}
Serpent.
\newblock \url{https://github.com/ethereum/serpent}, 2019.
\newblock [Online; accessed 01/09/2019].

\bibitem{vyper}
Vyper.
\newblock \url{https://github.com/ethereum/vyper}, 2019.
\newblock [Online; accessed 01/09/2019].

\bibitem{sample-trans}
Transaction 0x45725e88bcbfd30afacf184a58b1ea...
\newblock \url{https://tinyurl.com/y7tkoggh}, 2019.
\newblock [Online; accessed 01/09/2019].

\bibitem{batch-news}
Batchoverflow exploit creates trillions of ethereum tokens, major exchanges
  halt erc20 deposits.
\newblock \url{https://tinyurl.com/yb6eo6r9}, 2018.
\newblock [Online; accessed 01/09/2019].

\bibitem{racket}
The racket language.
\newblock \url{https://racket-lang.org/}, 2017.
\newblock [Online; accessed 01/09/2019].

\bibitem{non-interfere}
Non-interference (security).
\newblock \url{https://en.wikipedia.org/wiki/Non-interference_(security)},
  2018.
\newblock [Online; accessed 01/09/2019].

\bibitem{NiemetzPreinerBiere-JSAT15}
Aina Niemetz, Mathias Preiner, and Armin Biere.
\newblock Boolector 2.0 system description.
\newblock {\em Journal on Satisfiability, Boolean Modeling and Computation},
  9:53--58, 2014 (published 2015).

\bibitem{gasless}
Unchecked return values for low level calls.
\newblock \url{https://dasp.co/#item-4}, 2018.
\newblock [Online; accessed 01/09/2019].

\bibitem{CousotC77}
Patrick Cousot and Radhia Cousot.
\newblock Abstract interpretation: {A} unified lattice model for static
  analysis of programs by construction or approximation of fixpoints.
\newblock In {\em Proc. Symposium on Principles of Programming Languages},
  pages 238--252, 1977.

\bibitem{smart-sec}
Ethereum smart contract security best practices.
\newblock \url{https://consensys.github.io/smart-contract-best-practices/},
  2018.
\newblock [Online; accessed 01/09/2019].

\bibitem{mythril}
Mythril classic.
\newblock \url{https://github.com/ConsenSys/mythril-classic}, 2018.
\newblock [Online; accessed 12/01/2018].

\bibitem{fuzz-docker}
The ethereum smart contract fuzzer for security vulnerability detection.
\newblock \url{https://github.com/gongbell/ContractFuzzer}, 2018.
\newblock [Online; accessed 01/09/2019].

\bibitem{fuzz-data}
The ethereum smart contract fuzzer for security vulnerability detection.
\newblock \url{https://github.com/gongbell/ContractFuzzer}, 2018.
\newblock [Online; accessed 01/09/2019].

\bibitem{geth}
Official go implementation of the ethereum protocol.
\newblock \url{https://github.com/ethereum/go-ethereum}, 2018.
\newblock [Online; accessed 01/09/2019].

\bibitem{symbolic-e}
James~C King.
\newblock Symbolic execution and program testing.
\newblock {\em Communications of the ACM}, 19(7):385--394, 1976.

\bibitem{manticore}
Manticore.
\newblock \url{https://github.com/trailofbits/manticore/}, 2016.
\newblock [Online; accessed 01/09/2019].

\bibitem{ecf}
Shelly Grossman, Ittai Abraham, Guy Golan{-}Gueta, Yan Michalevsky, Noam
  Rinetzky, Mooly Sagiv, and Yoni Zohar.
\newblock Online detection of effectively callback free objects with
  applications to smart contracts.
\newblock In {\em Proc. Symposium on Principles of Programming Languages},
  pages 48:1--48:28, 2018.

\bibitem{smack}
Zvonimir Rakamaric and Michael Emmi.
\newblock {SMACK:} decoupling source language details from verifier
  implementations.
\newblock In {\em Proc. International Conference on Computer Aided
  Verification}, pages 106--113, 2014.

\bibitem{Hirai17}
Yoichi Hirai.
\newblock Defining the ethereum virtual machine for interactive theorem
  provers.
\newblock In {\em Financial Cryptography and Data Security - {FC} 2017
  International Workshops, WAHC, BITCOIN, VOTING, WTSC, and TA, Sliema, Malta,
  April 7, 2017, Revised Selected Papers}, pages 520--535, 2017.

\bibitem{GrishchenkoMS18}
Ilya Grishchenko, Matteo Maffei, and Clara Schneidewind.
\newblock A semantic framework for the security analysis of ethereum smart
  contracts.
\newblock In {\em Principles of Security and Trust - 7th International
  Conference, {POST} 2018, Held as Part of the European Joint Conferences on
  Theory and Practice of Software, {ETAPS} 2018, Thessaloniki, Greece, April
  14-20, 2018, Proceedings}, pages 243--269, 2018.

\bibitem{coq}
The coq proof assistant.
\newblock \url{https://coq.inria.fr/}, 2016.
\newblock [Online; accessed 01/09/2019].

\bibitem{isabelle}
Isabelle.
\newblock \url{https://isabelle.in.tum.de/}, 2016.
\newblock [Online; accessed 01/09/2019].

\bibitem{porosity}
Porosity.
\newblock \url{https://github.com/comaeio/porosity/}, 2016.
\newblock [Online; accessed 01/09/2019].

\bibitem{Erays}
Yi~Zhou, Deepak Kumar, Surya Bakshi, Joshua Mason, Andrew Miller, and Michael
  Bailey.
\newblock Erays: Reverse engineering ethereum's opaque smart contracts.
\newblock In {\em Proc. USENIX Security Symposium}, pages 1371--1385, 2018.

\bibitem{AvgerinosCHB11}
Thanassis Avgerinos, Sang~Kil Cha, Brent Lim~Tze Hao, and David Brumley.
\newblock {AEG:} automatic exploit generation.
\newblock In {\em Proc. The Network and Distributed System Security Symposium},
  2011.

\bibitem{ChaARB12}
Sang~Kil Cha, Thanassis Avgerinos, Alexandre Rebert, and David Brumley.
\newblock Unleashing mayhem on binary code.
\newblock In {\em Proc. IEEE Symposium on Security and Privacy}, pages
  380--394, 2012.

\end{thebibliography}

\end{document}